%% file: phasemod.tex
\documentclass{iopart}

\newcommand{\kbar}{\mathchar'26\mkern-9mu k}

\usepackage{graphicx}
\usepackage{tikz}
\usepackage{amssymb}
\usepackage{amstext}

\definecolor{glass}{HTML}{A4EBF3}
\definecolor{theGrey}{HTML}{E2E1D3}
\definecolor{green1}{HTML}{A9DC53}
\definecolor{orange1}{HTML}{F9AB60}
\definecolor{gold1}{HTML}{F5D85C}
\definecolor{pink1}{HTML}{F4678c}
\definecolor{purple1}{HTML}{D19DD4} 

\begin{document}
\title{Phase Noise in the Delta Kicked Rotor: From Quantum to Classical}
\author{D H White, S K Ruddell and M D Hoogerland}
\address{Department of Physics, University of Auckland, Private Bag 92019, Auckland, New Zealand}
\ead{donald.white@auckland.ac.nz}

\begin{abstract}
We experimentally investigate the effects of phase noise on the resonant and non-resonant dynamics of the atom-optics kicked rotor. Employing sinusoidal phase modulation at various frequencies, resonances are found corresponding to periodic phase shifts, resulting in the effective transformation of quantum anti-resonances into resonances and vice-versa. The stability of the resonance is analysed, with the aid of experiments, $\epsilon$-classical theory and numerical simulations, and is found to be surprisingly robust against phase noise. Finally we look into the effects of phase noise on dynamical localization and discuss the destruction of the localization in terms of decoherence.
\end{abstract}
\maketitle

\section{Introduction}

Noise and decoherence effects are of great importance to quantum systems. The `quantum' nature of the dynamics of a system relies heavily on the precise phase relationships between the degrees of freedom. Small levels of noise disrupt these correlations, leading to more `classical' behaviour~\cite{Zurek2003}. In any practical implementation of quantum systems, such as a quantum computer, the need to maintain high degrees of coherence means that noise is one of the major limiting factors of these devices. A tool to fully investigate noise effects in quantum systems is therefore required. The $\delta$-kicked rotor is one such tool, offering a good platform from which to examine the effects of noise on a quantum system, owing to the high degrees of control available to the experimenter. 

The $\delta$-kicked rotor system has long been a paradigm for studying quantum chaos. Since the seminal work by the Raizen group using cold atoms~\cite{Moore1994,Moore1995}, quantum chaotic systems have become accessible to experiment in a highly controllable way. Experiments utilising Bose-Einstein condensates (BECs) allow for precise control over the initial condition, which can span much less than the full momentum phase space. The unique quantum properties of BECs have been exploited to demonstrate, for example, the sensitivity of the kicked rotor to the quasi-momentum~\cite{Currivan2009}, the accelerator mode phase space~\cite{Behinaein2006} and time reversal of a classically chaotic system~\cite{Ullah2011}.

Modulation effects on the kicked rotor have generated much interest in recent years. Delande and co-workers employed amplitude modulation, with three incommensurate frequencies, to demonstrate an effective metal-insulator transition~\cite{Chabe2008}. This arose from an effective mapping to a three-dimensional Anderson model, which showcased dynamical localization in a new light. The modulated kicked rotor has also been used to demonstrate simple factorization~\cite{Sadgrove2008}, a quantum ratchet~\cite{White2013}, quantum enhancement of momentum transport~\cite{Sadgrove2013} and the effects of amplitude and kick frequency noise on the quantum resonance~\cite{Sadgrove2005a}.

The influence of noise on the kick frequency has been studied, demonstrating the destruction of dynamical localization and of the quantum resonance~\cite{Ammann1998}. In contrast, these features have been shown to be remarkably stable against amplitude noise~\cite{Sadgrove2004}. It is then an interesting question to look into the effects of phase noise. A previous investigation found that small amounts of spontaneous emission are analogous to phase noise, with experimental results revealing a striking enhancement of the energy at resonance when subject to spontaneous emission~\cite{Darcy2001}.

In this paper, we investigate more deeply the effects of phase noise in the delta-kicked rotor system. We consider the influence of sinusoidal phase modulation on both the resonant and non-resonant dynamics of the kicked rotor. We find resonant modulation frequencies at certain rational multiples of the kicking frequency, owing to the induction of periodic phase shifts of the applied potential.

Furthermore, we look into the effects of incommensurate phase modulation frequencies on the kicked rotor dynamics. The incommensurate frequency results in a pseudo-random phase sequence, which gives us repeatability and control for phase noise. In conjunction with numerical and $\epsilon$-classical analysis, we find that the quantum resonances are remarkably robust to phase noise. This is surprising given that the origin of the resonances lies in quantum interference --- ostensibly strongly dependent on precise phase relationships. We also explore the long-time behaviour, investigating the effect of noise on dynamical localization and highlighting the destruction of dynamical localization.

\section{Theoretical background}
\label{sec:theory}

\subsection{The $\delta$-kicked rotor}

In the classical $\delta$-kicked rotor model, a pendulum is subject to a pulsed gravitational field. The pulses are modelled as instantaneous Dirac-delta functions. At other times, the pendulum is free to rotate in an environment free of any external forces. The kicked rotor exhibits chaotic behaviour for sufficiently large kick-strengths~\cite{Chirikov1979}. The Hamiltonian for this system (in scaled, dimensionless units) is

\begin{equation}
	H(t) = \frac{\kbar p^2}{2}+k\cos(x)\sum_{i=0}^{N-1} \delta(t-i),
\label{eq:hamiltonian_classical}
\end{equation}

\noindent where $\kbar$ is the pulse period, $p$ is the angular momentum of the pendulum, $k$ is the kick-strength, $N$ is the number of kicks and $x$ is the angular position.

The classical kicked rotor was extended to the quantum realm and brought to experimental realisation through atom optics~\cite{Raizen1999}. This allowed access to a quantum chaotic system in a direct way. In the atom optics system, the external force is not gravity, but a pulse of an optical standing wave. The standing wave is created from the interference of two counterpropagating laser beams, detuned from the atomic resonance. This results in the atoms experiencing a sinusoidal phase grating. 

Interesting dynamics emerge from this system. The `chaos' inherent in the classical system is not strictly present in the quantum system owing to the linearity of the equations of quantum mechanics~\cite{Nakamura1993}. Instead, the chaos manifests as dynamical localization. This curious phenomenon has strong links with Anderson localization~\cite{Fishman1982}. The effect is that, after a short period where apparent classical motion occurs (termed the `quantum break time'), diffusion in momentum space is inhibited and the system reaches a steady state where no further energy is delivered to the system.

Another specifically quantum feature of the behaviour surrounds the appearance of a `quantum resonance'. For certain kicking frequencies ($\omega_k = 4\omega_r$, where $\omega_r$ is the atomic recoil frequency) the free evolution is completely negated, and subsequent kicks continue to add in phase constructively. This leads to ballistic energy growth. The origin of this behaviour is similar to the Talbot effect in near-field diffraction optics, where after certain propagation distances, the diffraction grating is re-imaged. In the quantum resonance case, the `re-imaging' occurs after a certain propagation time, known as the `Talbot time', $T_{tal}$.

Another important concept is that of the `anti-resonance'. At half-integer multiples of the Talbot time, the effect of the free evolution is for each successive kick to add destructively. Thus, the energy simply oscillates in time.

\subsection{Phase Modulation}

Our system is an extension of the classic system of (\ref{eq:hamiltonian_classical}), with the addition of a sinusoidal phase modulation of the standing wave phase grating.  Our system is now described by the time-dependent Hamiltonian (again in scaled, dimensionless units):

\begin{equation}
	H(t) = \frac{\kbar p^2}{2}+k\cos[x + \alpha\cos(\omega_p t)]\sum_{i=0}^{N-1} \delta(t-i),
\label{eq:hamiltonian}
\end{equation}

\noindent with phase modulation frequency $\omega_p$ and phase modulation amplitude $\alpha$. The momentum co-ordinate $p$ now represents the linear momentum of the atoms, while $x$ is the linear position co-ordinate. Definitions of these units in terms of experimental parameters can be found in Section~\ref{sec:kickParameters}.

In the classical system of a rotating pendulum~\cite{Nakamura1993}, such phase modulation would correspond to the direction of gravity rotating with frequency $\omega_p$. In our system, we modulate the phase of one of two interfering laser beams, thereby modulating the position of the optical grating.

\subsection{Basic concepts}
\label{sec:preliminary}

\begin{figure}[t]
\centering
\includegraphics[width=300px]{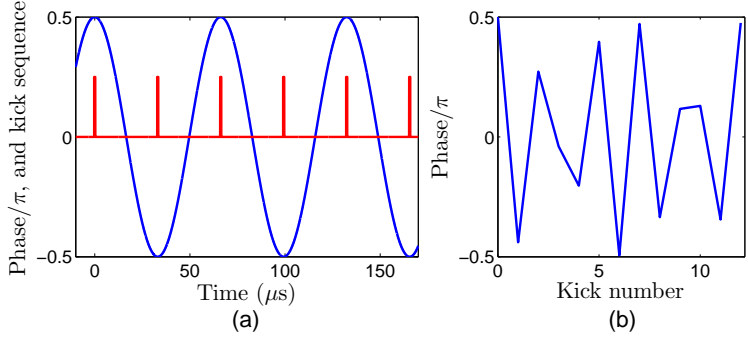}
\caption{In (a), illustration of the effect of phase modulating at 15 kHz, whilst kicking at 30 kHz (at the 33.1 $\mu$s anti-resonance). The red plot represents the timing of the kicks (the vertical scale is not relevant). Note the phase jump of $\pi$ from kick to kick. In (b), illustration of the effect of phase modulation at an incommensurate frequency. Here, $\omega_p=\sqrt{3}\omega_k/4$ is chosen as an example to indicate the pseudo-random nature of the kick sequence that arises from such a choice of $\omega_p$.}
\label{fig1}
\end{figure}

Many of the key effects of phase modulation on the kicked rotor dynamics can be understood with the aid of a simple picture, such as that in Figure~\ref{fig1}(a). Here, we consider kicking with pulse period $T_{tal}/2$ ($T_{tal}$ is the Talbot time of 66.3 $\mu$s for $^{87}$Rb). The phase modulation frequency is set to half the kicking frequency: $f = 1/T_{tal} \approx 15$~kHz. As the duration of a kick is far shorter than the timescale over which the phase changes appreciably, the phase sequence effectively reduces to a Nyquist sampling problem. It is straightforward to see that, if the modulation amplitude is set to $\pi/2$, the phase will jump by $\pi$ from kick to kick. This has the effect of completely negating the quantum anti-resonance, and effectively creating a quantum resonance condition where the energy grows ballistically with kick number.

Other pertinent points can be gathered from similar pictures. For example, phase modulation at the kicking frequency of $\approx 30$~kHz is equivalent to zero phase modulation, as the phase is the same for each kick. Even higher kicking frequencies are simply aliased to within the 0-15 kHz range (e.g.,~45 kHz is equivalent to 15 kHz). These stem from the fact that the Nyquist frequency for the system is 15 kHz. Another important aspect concerns the phase at which the kick sequence begins. If, for example, kicking occurs at 15 kHz and the phase goes as $\sin(\omega_p t)$ rather than $\cos(\omega_p t)$, then the phase will be constant for each kick. It is therefore important that the initial phase is controlled for the experiment.

\subsection{Analysis}

We can develop a more quantitative understanding of the system by considering the spectral qualities of the kick sequence. A train of $\delta$-functions in time has a Fourier spectrum of a comb of frequencies, spaced at the kicking frequency:

\begin{equation}
\mathcal{F}\left[\sum_{n=-\infty}^{\infty} \delta(t-nT)\right] = \sum_{m=-\infty}^{\infty} \delta(\omega-m\omega_k).
\end{equation}

Now we consider the phase modulated problem. The phase shift can be built in to the kick sequence, so we consider the Fourier transform:

\begin{eqnarray}
	F_{mod}(\omega) &= \mathcal{F}\left[\sum_{n=-\infty}^{\infty}\delta(t-nT)e^{i\alpha\cos\omega_pt}\right]  \\ &=\mathcal{F}\left[\sum_{n=-\infty}^{\infty}\delta(t-nT)\right]\ast\mathcal{F}\left(e^{i\alpha\cos\omega_pt}\right).
\end{eqnarray}

The exponential may be expanded in terms of Bessel functions~\cite{Abramowitz1965} to give the simple expression for the Fourier transform of:

\begin{eqnarray}
\mathcal{F}\left(e^{i\alpha\cos\omega_pt}\right) &= \mathcal{F}\left[J_0(\alpha)+2\sum_{k=1}^{\infty}i^k J_{k}(\alpha)\cos(k\omega_pt)\right] 
\\ &= 2\pi \sum_{k=-\infty}^{\infty} i^k J_k(\alpha)\delta(\omega-k\omega_p).
\label{eq:fourierSpec}
\end{eqnarray}

The full spectrum $F_{mod}$ is then the convolution of (\ref{eq:fourierSpec}) with a train of $\delta$-functions, which simply results in a continual repeating sequence of the spectrum in (\ref{eq:fourierSpec}).

Based on the simple picture introduced in Section~\ref{sec:preliminary}, we now make the heuristic assumption that the induced resonance is due only to the frequency component at $\omega_p = \omega_k/2$ and higher frequencies which alias to it. The spectrum in (\ref{eq:fourierSpec}) indicates that frequency components of integer multiples of $\omega_p$ are produced, with Bessel function weightings in the modulation strength $\alpha$. We then simply look at the fraction of the power spectrum concentrated at $\omega_k/2$ to estimate the extent of the influence of the modulation on the rotor energy.

The Fourier analysis also permits a prediction of the frequencies at which a resonance can be induced. From (\ref{eq:fourierSpec}), integer multiples of $\omega_p$ are produced. Odd multiples of $\omega_k/2$ are aliased to $\omega_k/2$. This gives the simple resonance condition:

\begin{equation}
\omega = \frac{2n+1}{m}\frac{\omega_k}{2}, \qquad n, m \in \mathbb N.
\end{equation}

\begin{figure}[t]
\centering
\includegraphics[width=250px]{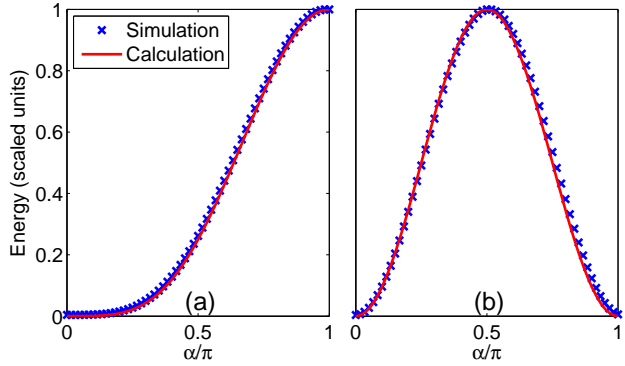}
\caption{Overlay of quantum simulations, conducted through Floquet operations on an initial wavepacket, and calculations based on the Fourier spectrum in (\ref{eq:fourierSpec}). (a) 7.5 kHz modulation with summation in (\ref{eq:calc1}) and (b) 15 kHz modulation with summation in (\ref{eq:calc1}). The energy has been scaled to a maximum value of 1. To prevent quasimomentum effects from complicating the issue, the simulations have been conducted using a narrow Gaussian initial condition, $\psi(p) = \exp(-p^2/2\sigma^2)$, where $\sigma = 0.05p_r$. The simulations are over 14 kicks, with $\kbar = 2\pi$.}
\label{fig:besseltest}
\end{figure}

\noindent We now illustrate the applicability of the Fourier spectrum in predicting the energy as a function of modulation strength. For the first order resonance, the relevant harmonics are the $1^{st}$ (15 kHz), $3^{rd}$ (45 kHz), $5^{th}$ (75 kHz), $\ldots$, while for the second order resonance, the relevant harmonics are the $2^{nd}$, $6^{th}$, $10^{th}$, $\ldots$ The relevant harmonics for an $m^{th}$ order resonance are clearly $m(2n+1), n \in \mathbb Z$. We look at the fraction of the power spectrum concentrated at these harmonics, to give a general rule for the energy of an $m^{th}$ order resonance as

\begin{equation}
E_m(\alpha) \propto \left|\sum_{n=0}^{\infty} i^{m(2n+1)}J_{m(2n+1)}(\alpha)\right|^2
\label{eq:calc1}.
\end{equation} 

\noindent In Figure~\ref{fig:besseltest}, we compare these calculations with quantum simulations (explained in Section~\ref{sec:numerics}). The simulations have been conducted directly using the Floquet operator and are independent of the Fourier analysis. Both the 7.5 kHz and the 15 kHz resonances are plotted as a function of modulation strength $\alpha$. The correspondence between the two is practically perfect, indicating the validity of the Fourier treatment.

\subsubsection{Incommensurate frequency}

~\\Another simple application of the Fourier treatment gives a different insight into why an incommensurate frequency can be regarded as representing a true `noisy' signal. From (\ref{eq:fourierSpec}), harmonics of the modulation frequency are produced. If the modulation frequency is incommensurate with the kicking frequency, then the higher harmonics are aliased back to a wide range within the 0-$\omega_k/2$ region. A histogram reveals that all frequencies within this range are close to equally populated: an incommensurate frequency results in true `white noise'. (This argument is strictly only true when the number of kicks is large, as a finite $N$ causes the higher harmonics to be diminished. However, the idea still holds.)

\subsection{Phase Noise}

An interesting effect occurs if $\omega_p$ is chosen to be incommensurate with the kicking frequency. As indicated in Figure~\ref{fig1}(b), the phase sequence for such a choice of frequency is pseudo-random. It is an interesting question to ask what effect this `phase noise' will have on dynamical localization.

The original, unmodulated kicked rotor is a very well known classically chaotic system. In the quantum limit, this chaos manifests as dynamical localization for pulse periods incommensurate with the quantum resonances (see, for example,~\cite{Nakamura1993}). This is due to the direct mapping of the Hamiltonian onto the Anderson model~\cite{Grempel1984}. In a similar vein, when the kicked rotor potential is a function of $N$ pairwise incommensurate frequencies, the system maps to an $N-1$ dimensional Anderson problem~\cite{Casati1989}. This has been experimentally verified by Delande and co-workers, with three incommensurate frequencies controlling amplitude modulation of the kick pulses~\cite{Chabe2008}.

The previous works suggest that, in our system, kicking on resonance with an incommensurate phase modulation frequency would result in a one-dimensional Anderson problem, with dynamical localization eventuating after a quantum break time. Kicking off resonance, with an addition of a further incommensurate $\omega_p$, would result in a two dimensional Anderson problem, similar to that investigated by Doron and Fishman~\cite{Doron1988}. The localization length $\xi$ in this case grows should grow exponentially with the ``mean free path", which is a function of the kick strength. As suggested in the scaling theory of localization,  all eigenstates are localized in a two dimensional system, meaning localization should eventually result with large $\xi$~\cite{Abrahams1979}.

\subsection{$\epsilon$-classical theory}

Our analysis in the vicinity of the resonances makes use of $\epsilon$-classical theory, developed in~\cite{Wimberger2004}. The method revolves around making a pseudo-classical approximation to the dynamics in the vicinity of the quantum resonance. We write the pulse period $\kbar = 2\pi\ell+\epsilon$, where $\ell$ is the resonance order. An important parameter is $\epsilon$, which represents the detuning from quantum resonance. $|\epsilon|$ takes the role of the effective Planck's constant for our pseudo-classical analysis, even though $\kbar$ is large. By writing $J = \epsilon N + \pi\ell+\kbar\beta$ and $\vartheta=x+\pi[1-\text{sgn}(\epsilon)]/2$ (where $\beta$ is the quasimomentum and $N$ is the integer momentum), we arrive at the pseudo-classical map:

\begin{equation}
J_{t+1}=J_{t}+k|\epsilon|\sin(\vartheta_{t+1}),\qquad\vartheta_{t+1}=\vartheta_t+J_{t+1}.
\label{eq:pseudoClassical}
\end{equation}

The main importance in the map lies in the resultant effective stochasticity parameter, $k|\epsilon|$: a scaling law is obeyed for kick-strength $k$ and resonance detuning $|\epsilon|$. The map in (\ref{eq:pseudoClassical}) will be used for the analysis of the resonance stability in Section~\ref{sec:robustResonance}.

\section{Methods}

\subsection{Experimental setup}

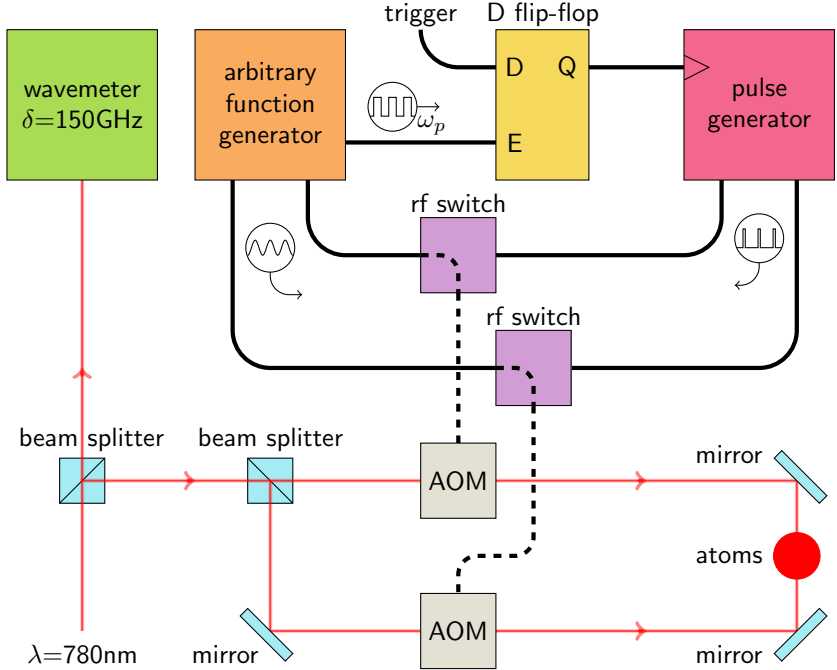
\begin{figure}[t]
\centering
\input{figure3.tex}
\caption{Schematic of experimental setup. Interference of the two linearly polarised detuned beams creates a standing wave. The dual channel arbitrary function generator (AFG) generates two 80 MHz signals to drive acousto-optic modulators (AOMs), with one channel phase modulated at the desired frequency. The TTL output of the AFG is used to trigger a D flip-flop, which ensures that the initial phase is synchronised with the start of the kicking sequence.}
\label{fig_setup}
\end{figure}

We begin with a $^{87}$Rb Bose-Einstein condensate (BEC) of approximately 20 000 atoms, prepared in an all-optical BEC machine~\cite{Wenas2008}. The setup is similar to that of our previous phase modulated kicked rotor work~\cite{White2013}. 

\subsubsection{Kick parameters}
\label{sec:kickParameters}

~\\As indicated in Figure~\ref{fig_setup}, an optical standing wave is created via interference of two counter-propagating laser beams overlapping the atoms, detuned by $\Delta/(2\pi)=150$~GHz from the $^{87}$Rb D2 line. The AC Stark shift means that the standing wave functions as a phase grating. Acousto-optic modulators (AOMs) provide individual control over each of the two beams. The AOMs are driven with a dual channel Tektronix arbitrary function generator (AFG3252), running at 80 MHz, and switched with Minicircuits rf switches (ZSWA-4-30DR). The AOMs present us with the ability to control the timing and relative phase of the two beams, giving us command over the phase grating.

A single kick results from a $\tau=300$~ns pulse of the optical lattice. The short ``on" period, with respect to the atomic velocity and the standing wave period of $\lambda/2$, means that the diffraction takes place firmly in the Raman-Nath regime. This allows us to mathematically treat each kick as a $\delta$-function. Discrete momentum transfers of 2 photon recoils ($2\hbar k_L = 2p_r$) are allowed in this system, as absorption occurs in one beam and stimulated emission in the other.

For convenience, we utilise the following scaled units throughout:

\begin{eqnarray}
x = 2k_LX, \\
p = \frac{P}{2\hbar k_L}, \\
t = t^{\prime}/T, \\
k = \frac{\tau \Omega^2}{4\Delta}, \\
\kbar = \frac{4\hbar k_L^2 T}{m}, \\
\epsilon = \kbar - 2\pi\ell, \qquad \ell \in \mathbb Z, |\epsilon| < \pi
\end{eqnarray}

\noindent where $X$ and $P$ are real position and momentum operators; $k_L = 2\pi/\lambda$ is the wavenumber of the kicking laser; $T = 2\pi/\omega_k$ is the kick period; $t^{\prime}$ is real time; $m$ is the atomic mass; $\tau$ is the duration of a single kicking pulse; $\Delta$ is the laser detuning from resonance; and $\Omega$ is the Rabi frequency.

The pulses are spaced at intervals near either the resonant Talbot time of 66.3 $\mu$s, or the anti-resonant time of 33.1 $\mu$s. The timing is controlled with a programmable pulse generator. As mentioned in Section~\ref{sec:theory}, we require the modulation to be cosinusoidal with respect to the first kick. We therefore need to synchronise the start of the kick sequence with the phase modulation of the function generator. This is accomplished through the use of the logic output of the function generator. The pulse generator trigger signal is delayed by a D flip-flop, until the flip-flop is triggered by the rising edge of the TTL signal. A further delay of one quarter of the phase modulation period is programmed in to the start of the pulse sequence to ensure that our phase modulation is $\cos(\omega_pt)$, rather than $\sin(\omega_pt)$.

\subsubsection{Analysis}

~\\The atoms are allowed to freely expand for 3 ms following the pulse sequence and are imaged using an absorption technique. Repump light is delivered for 100 $\mu$s to pump the atoms into the hyperfine $F=2$ state before detection. A probe beam resonant with the $F=2\rightarrow F=3$ transition is then applied and is imaged with a 4f system. This time-of-flight procedure effectively maps out the momentum distribution of the atoms.

The data analysis begins with removal of background image noise via a least-squares fitting procedure. The relevant slice of the image is cropped, and summed across to give a one dimensional momentum profile. We extract the energy ($E/E_r=\langle (p/p_r)^2\rangle$) by finding the numerical variance of the 1D momentum profile; we extract the zero momentum population by finding the proportion of atoms within $|p| < 1\hbar k_L$.

\subsection{Numerical methods}
\label{sec:numerics}

We conduct numerical simulations in order to corroborate our experimental data with theoretical results. The simulations are conducted in one dimension using the split-step method, which is ideally suited to the $\delta$-kicked rotor problem. The evolution can be directly described by the Floquet operator:

\begin{eqnarray}
F(k,\omega_p,\kbar,\alpha,t) &= F_{free}F_{kick} \\
&= \exp\left(-i\frac{\kbar p^2}{2}\right)\exp\left[-ik\cos(x+\alpha\cos\omega_p t)\right].
\label{eq:floquet}
\end{eqnarray}

\noindent The Floquet operator $F$ describes the evolution from immediately prior to one kick to immediately prior to the next. The Floquet operator is useful as it can be separated into two distinct parts: the kick operator $F_{kick}$, and the free evolution operator $F_{free}$. The kick operator is diagonal in position space; the free evolution is diagonal in momentum space. A full kick, including free evolution, can then be completed with two multiplications and a single Fourier transform from position to momentum space.

Except where stated, we choose an initial condition of the solution to the Gross-Pitaevskii equation for our trap, which has been subject to 300 $\mu$s of expansion under mean field repulsion. This results in a wavepacket with a full-width at half-maximum of 0.4 $\hbar k_L$ in momentum space.

\section{Results}

\subsection{Effect of phase modulation frequency on quantum resonance}

As mentioned in Section~\ref{sec:theory}, the most dramatic effect of the phase modulation frequency on the dynamics relates to the transformation of an anti-resonance into a resonance, and vice versa, when the phase modulation frequency ($\omega_p$) is half of the kicking frequency ($\omega_k$). In this section we examine the nature of the resonance.

\begin{figure}[t]
\centering
\includegraphics[width=300px]{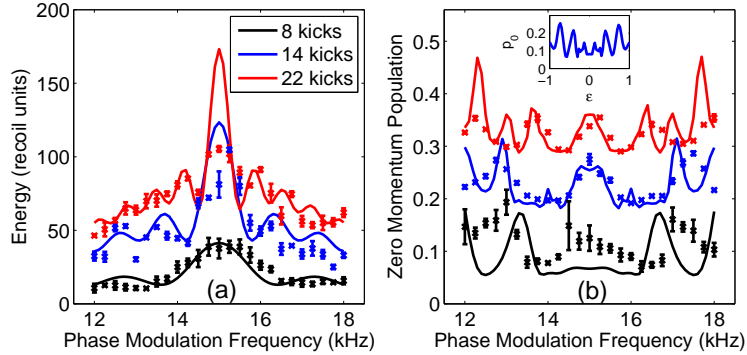}
\caption{Around 15 kHz resonance, with $\alpha=\pi/6$. (a) Experimental energies for 8, 14 and 22 kicks are overlaid with simulations. The kick strength $k$ has been calibrated to be 1.7, 2.0 and 2.1 for the 8, 14 and 22 kick data respectively. Energies are in recoil units: $\langle (p/p_r)^2\rangle$. (b) The population of the zero momentum state is overlaid with simulations. For clarity, the 14 kick data has been vertically offset by 0.15, and the 22 kick data has been offset by 0.25. (Inset) For comparison, a simulation of the zero momentum population ($p_0$) for 14 kicks, for no phase modulation, around the 66.3 $\mu$s resonance. The simulation is plotted as a function of $\epsilon$ (the pulse period detuning from resonance).}
\label{fig:15khz}
\end{figure}

In Figure~\ref{fig:15khz} we examine the behaviour in the vicinity of the 15 kHz resonance. Data is shown for 8, 14 and 22 kicks and is overlaid with simulations. The sinc-type resonance feature in the energy, indicated in the simulation, is well replicated by the data. The Fourier-based analysis in Section~\ref{sec:theory} suggests a Fourier $1/N$ scaling of the resonance width, and we observe this in both simulations and data. 

For the energy measurement, the main discrepancy between the data and simulations occurs directly on resonance. The high energies on resonance result in a number of weakly populated high momentum states, which contribute significantly to the energy. With our absorption imaging system, these are inefficiently detected, leading to lower than expected energies. Small imperfections also arise from the time-of-flight imaging method, which does not produce a perfect momentum map for short expansion times.

An interesting feature of the data is the significant peak in the zero momentum, shown in Figure~\ref{fig:15khz}(b). We include all atoms in the zeroth diffraction order in this measurement. This peak is rather counter-intuitive, as it could be expected that alongside the energy growth on resonance, fewer atoms would remain in the zero momentum state. Indeed, this prediction holds for the unmodulated kicked rotor resonances, as shown in the inset. One possible reason is the generation of a small ratchet effect. For a small number of kicks, the phase sequence may appear periodic, thereby generating small transporting island structures in phase space~\cite{White2013,Sadgrove2013}. This would lead to directed momentum transport --- hence a small peak in the zero momentum state.

\begin{figure}[t]
\centering
\includegraphics[width=300px]{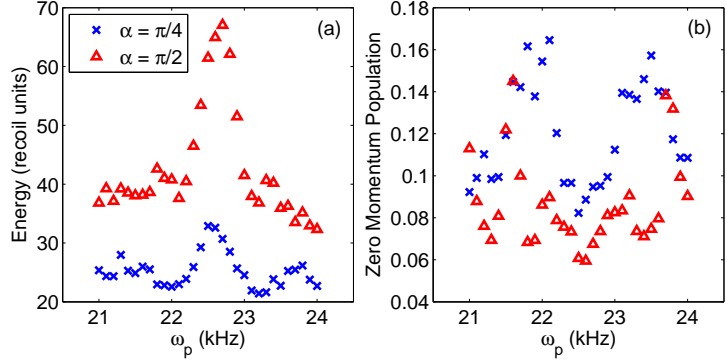}
\caption{Experimental data as a function of modulation frequency $\omega_p$ around the fractional resonance at 22.5 kHz. Different modulation amplitudes $\alpha$ are shown for energies in (a) and zero momentum populations in (b). Both datasets are for 14 kicks, $k = 2.0$.}
\label{fig:22khz}
\end{figure}

\begin{figure}[t]
\centering
\includegraphics[width=300px]{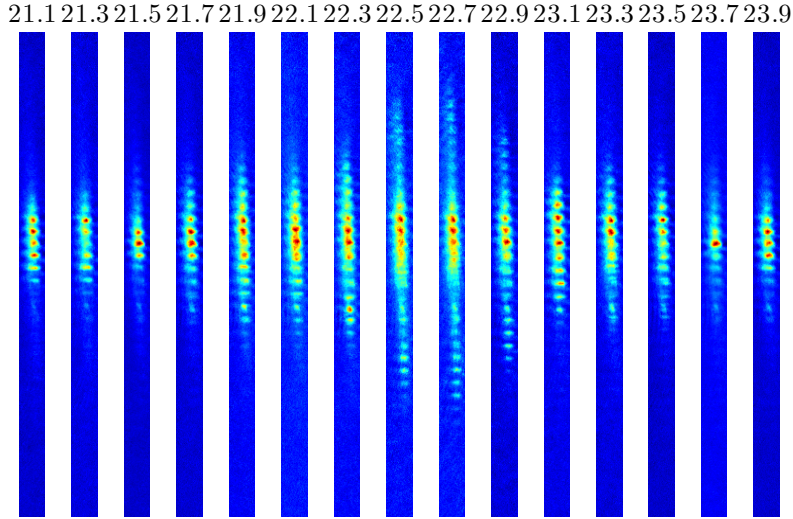}
\caption{Experimental images of the momentum distribution for different phase modulation frequencies. The data was collected with 14 kicks at 33.1 $\mu$s and $\alpha = \pi/2$. We measure the kick strength to be $k = 2$. The headings indicate the phase modulation frequencies in kHz. Data has been averaged over 4 experimental runs.}
\label{fig_resPic}
\end{figure}

In addition to the main resonance at 15 kHz, we observe narrow fractional resonances. In this system, fractional resonances occur due to the overlap of higher harmonics of $\omega_p$ with the resonant frequency $\omega_k/2$. The second order resonance at 22.5 kHz (equivalent to the resonance at 7.5 kHz), is shown in Figure~\ref{fig:22khz}. An interesting aspect of this resonance, confirmed by simulations, is the lack of a peak in the zero momentum population. Experimental images of the momentum distribution are shown in Figure~\ref{fig_resPic}.

\begin{figure}[t]
\centering
\includegraphics[width = 180px]{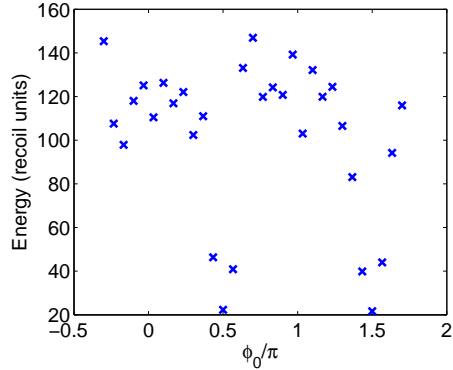}
\caption{Experimental data showing the influence of initial phase in modulation sequence on the kicked rotor energy. Clear troughs are shown for $\phi_0 = \pi/2$ and $3\pi/2$, corresponding to the modulation going as $\sin(\omega_pt)$. Data collected after 10 kicks, with $\alpha = \pi/2$, $\omega_p = 15$ kHz and $k$ = 2.0.}
\label{fig:initPhase}
\end{figure}

As previously mentioned, the initial phase $\phi_0$ of the modulation $\alpha\cos(\omega_pt+\phi_0)$ can have a dramatic effect. In Figure~\ref{fig:initPhase}, data is shown scanning over the initial phase. The strong dependence of the initial phase on the system meant that $\phi_0$ needed to be controllable: this was achieved through a flip-flop implementation (indicated in Figure~\ref{fig_setup}).

\begin{figure}[t]
\centering
\includegraphics[width=200px]{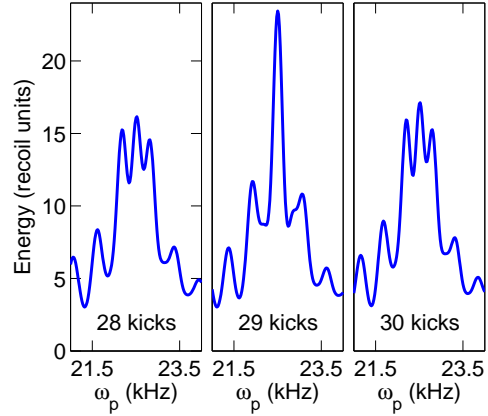}
\caption{Simulations indicating the effect of even or odd numbers of kicks on the second order phase modulation resonance. The resonance with 28, 29 and 30 kicks are shown. Simulation parameters: $k = 2$, $\ell = 1$, $\alpha = \pi/6$.}
\label{fig:evenOdd}
\end{figure}

Finally, it is worth noting the dependence of the resonance on even or odd numbers of kicks. The natural tendency of the unmodulated $\ell = 1$ anti-resonance is an oscillating energy, where each kick undoes the effect of the previous one. Odd numbers of kicks therefore produce greater energies than even numbers. For a kick sequence with a spectral component at $\omega_k$, an odd number of kicks will then provide increased energy on resonance. This leads to sharper resonances for odd $N$, as indicated in Figure~\ref{fig:evenOdd}.

\subsection{Phase Noise: Robustness of the quantum resonance}
\label{sec:robustResonance}

\begin{figure}[t]
\centering
\includegraphics[width=300px]{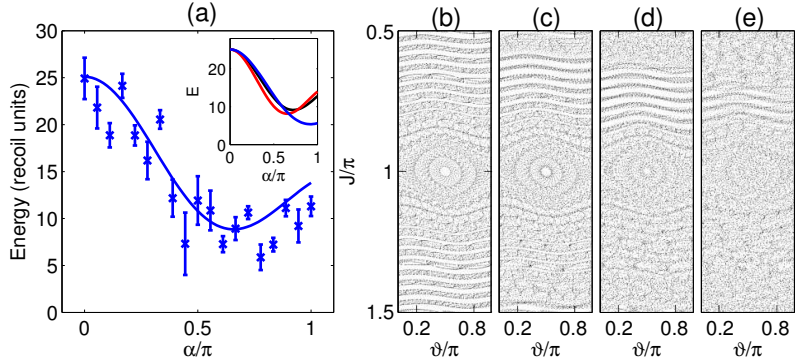}
\caption{(a) Experimental data and simulations indicating the fall-off from resonance with increasing phase `noise'. The data has been collected after 15 kicks on the main $\ell = 2$ resonance (pulse period 66.3 $\mu$s), with a phase modulation frequency of 6495 Hz and a kick strength of $k = 0.65$. (Inset) Similar simulations to that in (a), with different incommensurate phase modulation frequencies $\omega_k/\pi$ (black), $\omega_k/\sqrt{3}$ (red) and $\sqrt{5}\omega_k/3$ (blue). Poincar\'{e} plots are then shown for increasing phase noise, with $\omega_p = 6495$ Hz and $\alpha$ values (b) 0, (c) $\pi/18$, (d), $\pi/6$ and (e) $\pi/3$. Here $J$ is the momentum co-ordinate and $\vartheta$ is the position co-ordinate, and $k|\epsilon|$ is chosen to be 0.1.}
\label{fig:noise1}
\end{figure}

One nice advantage of using phase modulation is that it allows for replicable phase ``noise'' sequences to be produced. This means that the data does not need to be ensemble averaged over a number of noisy (yet random) sequences, in order for meaningful correlations between parameters and the noise level to be extracted. 

In Figure~\ref{fig:noise1}(a), data is presented with increasing levels of phase noise on the main 66.3 $\mu$s quantum resonance. The modulation frequency of 6495 Hz produces the same phase sequence as shown in Figure~\ref{fig1}(b). A decay in the energy is indicated with increasing $\alpha$. However, the decay rate is slow, and a substantial level of noise is required before the resonance is completely negated ($\alpha \approx  \pi/3$). Similar results are obtained for different incommensurate frequencies, shown in the inset, indicating that the microscopic noise details are not significant here.

The resonance stability indicated by the data and simulations is supported by the pseudo-classical Poincar\'{e} surfaces of section in Figures~\ref{fig:noise1}(b)-(e). The central resonance island becomes increasingly blurred by the phase noise, yet maintains its identity with very large phase noise amplitudes. This indicates that the resonance is surprisingly robust to phase noise.

\begin{figure}[t]
\centering
\includegraphics[width=200px]{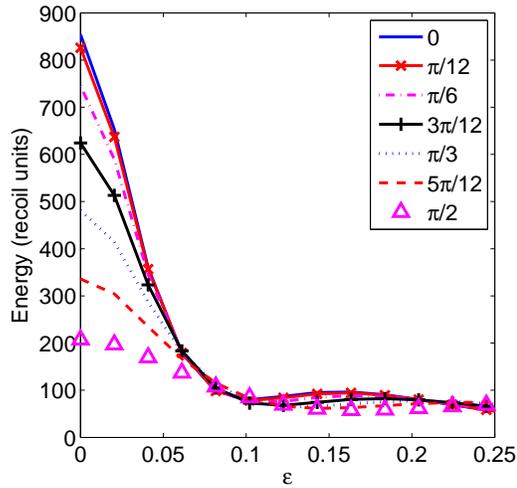}
\caption{Simulated pulse period energy resonance curves for increasing levels of phase noise. 30 kicks are performed around the $\ell = 2$ resonance, with $\omega_p = 6495$~Hz. The legend shows the different modulation amplitudes $\alpha$.}
\label{fig:noise2}
\end{figure} 

The robustness of the resonance is further supported by simulations. In Figure~\ref{fig:noise2}, energy is plotted as a function of $\epsilon$ around the $\ell = 2$ resonance ($\approx 66.3~\mu$s pulse period). Phase modulation is conducted at 6495 Hz. The simulations support the pseudo-classical phase space analysis in Figures~\ref{fig:noise1}(b)-(e), indicating the persistence of the resonance peak for large levels of phase noise.

\subsubsection{Relation to previous studies}

~\\Investigations have previously been carried out into the effects of amplitude noise on quantum resonances~\cite{Sadgrove2005a,Sadgrove2008b}. Amplitude noise affects the resonance behaviour due to the single-parameter dependence of the pseudo-classical map on $k|\epsilon|$~\cite{Wimberger2004,Sadgrove2008b}. If $k_i=k_0(1+R_i)$, where $R_i$ is a random variable, then the classical phase space will be affected and the resonance will be diminished. While the changes to the classical phase space are non-negligible, it was found to a first approximation that the resonance would be qualitatively unaffected by amplitude noise. We obtain similar results for phase noise. For a small, yet non-negligible, phase modulation amplitude of $\pi/12$ with an incommensurate frequency, the resonance is qualitatively unchanged (shown in Figure~\ref{fig:noise2}). Larger modulation amplitudes begin to have a more pronounced effect, broadening the resonance.

Contrary to Figures~\ref{fig:noise1} and~\ref{fig:noise2}, the results of d'Arcy {\it et. al.} indicated an {\it increase} in energy on resonance when subject to spontaneous emission~\cite{Darcy2001}. This occurred due to the momentum spread of their initial condition, which was a thermal laser-cooled cloud of atoms. On resonance, various momentum classes of rational multiples of $\hbar k_L$ are subject to energy oscillations --- i.e. an anti-resonance condition --- due to the effective phase imprint precisely oscillating in time. Phase noise, or spontaneous emission, destroys this effect to give enhanced energy. Our results indicate a slow decay in energy with phase noise due to the narrow momentum distribution of our BEC ($\Delta p \ll \hbar k_L$), which allows access to a single resonance class.

\subsection{Phase Noise --- on resonance}

It is quite clear that if $\epsilon = 0$, super-diffusion will persist for all time in the presence of amplitude noise. The phase accumulated from free evolution remains a multiple of $2\pi$, meaning each kick adds constructively. With phase noise on $\epsilon = 0$, the outcome is not immediately obvious. Here we investigate the effects of noise on resonance.

Our experimental results were inconclusive, with linear energy growth observed up to 25 kicks. Our experimental setup is not designed to measure large energies due to the relatively low number of atoms in our BEC, which results in a low signal-to-noise ratio for the energy-important high momentum states. To answer the phase modulation question, we therefore decided to employ numerical simulations.

\begin{figure}[t]
\centering
\includegraphics[width=250px]{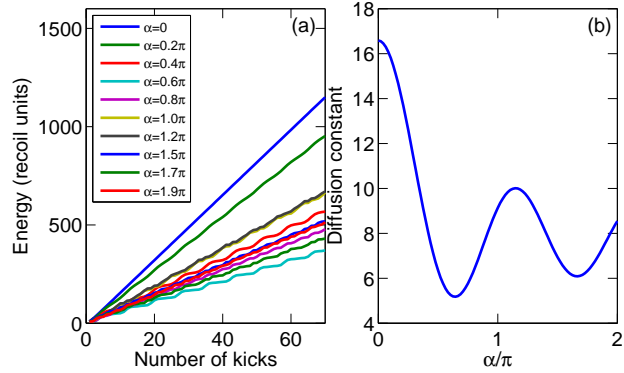}
\caption{Simulations indicating influence of phase noise on the quantum resonance. The kicking parameters are chosen to be $k = 2$, $\epsilon = 0$, $\ell = 2$, $\omega_p = \sqrt{3}\omega_k/4 = 6495$~Hz. (a) The energy as a function of time for various noise levels $\alpha$. (b) The diffusion constant (in units of energy recoils per kick) as a function of $\alpha$.}
\label{fig:noiseOnResonance}
\end{figure}

Our simulations in Figure~\ref{fig:noiseOnResonance} indicate approximate linear energy growth for all levels of phase noise\footnote{There are two things to note about Figure~\ref{fig:noiseOnResonance}. Firstly, the energy does not grow quadratically on resonance due to the spread in quasimomentum of the initial condition~\cite{Wimberger2005}. Secondly, the increases observed for $\alpha \gtrsim 2\pi/3$ are due to the maximum phase jump per kick now being larger than $\pi$.}. At first glance, this may appear surprising. Following the logic of Casati {\it et. al.}~\cite{Casati1989}, the presence of two incommensurate frequencies in the system may suggest a 1D Anderson system. If this was the case, the energy would stabilise after a `quantum break time', which is not observed. A more careful analysis reveals the reason for this behaviour.

When $\ell=2$ and $\epsilon = 0$, each momentum order accumulates a phase between kicks which is a multiple of $2\pi$: all momentum orders rephase. This means that the free evolution can, in a sense, be disregarded and the net effect is that of one `long' kick. To find the effective phase grating, we require a sum of the individual gratings:

\begin{equation}
V_{eff}(x,N) = \sum_{n=0}^{N-1}V_n(x) = \sum_{n=0}^{N-1}k\sin[x+\alpha\cos(n\omega_p T)],
\end{equation}

\noindent where $T$ is the pulse period and $N$ the number of kicks. With the individual gratings adding out of phase, the resultant amplitude is less than $Nk$. Performing this summation yields the same form of energy as a function of $\alpha$ as in Figure~\ref{fig:noiseOnResonance}(b). The `steps' in the time evolution seen in Figure~\ref{fig:noiseOnResonance}(a) are also found. Fundamentally, in this case, the resonance is not broken by phase noise, but is rather diminished due to the effective noise-induced reduction of kick-strength.

\begin{figure}[t]
\centering
\includegraphics[width=330px]{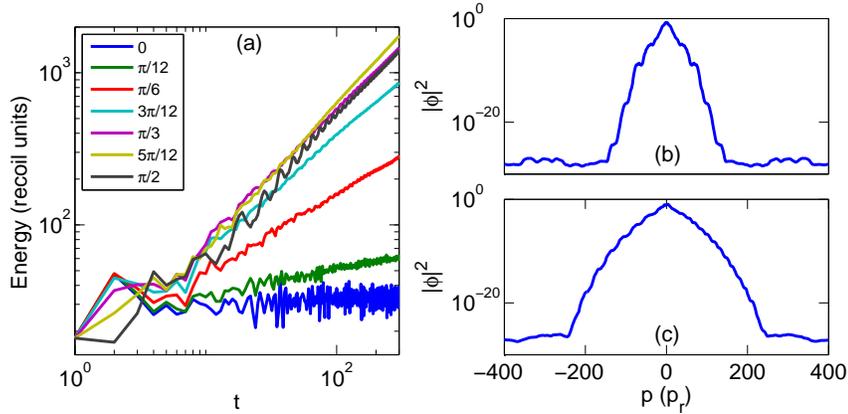}
\caption{Effect of phase noise on dynamical localization. (a) Simulations of energy as a function of number of kicks with increasing $\alpha$. $\omega_p = 6495$~Hz, $\epsilon = 0.4$, $k = 3$, $\ell = 2$. (b) Semilog plot of wavefunction in momentum space ($|\phi|^2$) for $\alpha = 0$ after 70 kicks. (c) Semilog plot of $|\phi|^2$ for $\alpha = \pi/6$ after 70 kicks. For zero modulation, strong dynamical localization is witnessed, evidenced by exponential localization. As the modulation strength is increased, the dynamical localization is destroyed.}
\label{fig:locsim}
\end{figure}

\subsection{Phase Noise: Effect on dynamical localization}

The next question we can ask is the effect that being away from quantum resonance will have ($\epsilon > 0$). We set $\kbar$, $\pi$ and $\omega_p$ to all be incommensurate. The value of $k|\epsilon| = 1.2$ is chosen to be larger than 0.97, such that the classical phase space is fully chaotic. Previous analyses suggest that a two-dimensional Anderson system would result from this system~\cite{Doron1988,Wang2014}.

In Figure~\ref{fig:locsim}(a) we plot simulations of energy as a function of time for different modulation strengths. In the case of zero modulation, the energy simply oscillates in time with no net growth: this is simple dynamical localization. As the noise level $\alpha$ is increased, the energy begins to grow with power-law behaviour, $E\propto t^q$. The exponent $q$ grows with $\alpha$ and tends to unity.

The noise-induced destruction of dynamical localization is well illustrated in Figures~\ref{fig:locsim}(b) and (c). The plots show the momentum space wavefunction after 70 kicks, (b) without and (c) with phase noise. The exponential localization in the absence of phase noise is clear in (b), with strong linearity in the semi-logarithmic plot. The nature of the wavefunction then changes markedly for $\alpha = \pi/6$, broadening the distribution and not giving any indication of exponential localization.

Dynamical localization relies on strict quantum correlations between momentum states. Small levels of phase noise disrupt these correlations, weakening the localization~\cite{Cohen1991}. Pure diffusion results when this coherence has been destroyed, resulting in classical-like behaviour. 

The contrast between the robustness of the resonance and the sensitivity of the localization highlights the fact that the transition from quantum to classical behaviour is not trivial. 

It is something of an open question as to whether it is meaningful to associate this with a 2D Anderson system --- the localization length would be extremely large (note clear diffusion for up to 300 kicks), and well beyond what could be experimentally observed. Physically, the relevant notion here relates to the inhibition of interference due to noise.

Based on the comparison between Figures~\ref{fig:noise1} and \ref{fig:noise2} and Figure~\ref{fig:locsim}, it is clear that dynamical localization is far more sensitive to noise effects than the quantum resonance.

\section{Conclusions}

We have experimentally and theoretically investigated the effects of phase modulation on the $\delta$-kicked rotor. Two main effects have been explored: the effect on quantum resonances, and the effective phase noise sequence induced through an incommensurate modulation frequency. 

Resonances have been found corresponding to rational multiples of $\omega_k/2$. These are induced through effective periodic phase shifts of the potential. We have heuristically derived a simple expression for the form of the kicked rotor energy as a function of modulation amplitude, which has been found to agree well with experiments and numerical simulations.

Incommensurate frequencies have been used to probe the effects of phase noise on the kicked rotor. The quantum resonance has been found to be surprisingly robust to phase noise: the resonance peak is significant even where phase jumps of up to $\pi$ per kick can be found. This is supported by a pseudo-classical map picture, which shows robust resonance islands. Further numerical analysis indicated the absence of dynamical localization for incommensurate frequencies when $\epsilon = 0$. Dynamical localization for finite $\epsilon$ was found to be sensitive to phase noise, with sub-diffusion witnessed for small levels of noise. The particular choice of incommensurate frequency was not found to have any significant impact on the results.

The presence of resonances for certain frequencies indicates the potential for `coloured' noise to strongly influence the dynamics of a quantum system, whilst remaining robust to white noise. Dynamical localization is, however, sensitive to noise both white and coloured.

Throughout, our experimental data was found to be in excellent agreement with numerical simulations. We hope this work will stimulate further interest in the modulated kicked rotor, with these experiments adding to a phalanx of previous intriguing dynamics discovered in modified kicked rotor systems.

\ack{The authors would like to thank Sandro Wimberger and Scott Parkins for helpful insights. We gratefully acknowledge the support of the Marsden Fund, with funding administered by the Royal Society of New Zealand.}

\section*{References}

\providecommand{\newblock}{}

\end{document}

%% file: figure3.tex


		\begin{tikzpicture}[font=\sffamily]

			\filldraw[draw=black,fill=glass](0.7,0.2)rectangle+(0.6,0.6);
			\draw(0.7,0.2)--(1.3,0.8);
			\node[align=center] at (1.12,1.05) {beam splitter};

			\filldraw[draw=black,fill=glass](3.2,0.2)rectangle+(0.6,0.6);
			\draw(3.2,0.8)--(3.8,0.2);
			\node at (3.5,1.05) {beam splitter};
			
			\draw[ultra thick,rounded corners=0.5cm](3,4.5)--(3,2.0)--(10.5,2.0)--(10.5,4.5);
			\draw[ultra thick,rounded corners=0.5cm](4,4.5)--(4,3.5)--(9.5,3.5)--(9.5,4.5);
			
			\draw[ultra thick,rounded corners=0.5cm](5.5,6.5)--(5.5,6)--(9,6);
			\node at (5.5,6.7){trigger};

			\draw[ultra thick,dashed](6,3)--(6,1);
			\draw[ultra thick,rounded corners=0.25cm,dashed](7,1.5)--(7,-0.5)--(6,-0.5)--(6,-1);

			\draw[red,ultra thick,draw opacity=0.2](1,0.5)--(1,4.5);
			\draw[red,draw opacity=0.5](1,0.5)--(1,4.5);
			\draw[red,thick,draw opacity=0.5](1,0.5)--(1,4.5);
			
			\draw[red,ultra thick,draw opacity=0.2](3.5,0.5)--(3.5,-1.5)--(10.5,-1.5)--(10.5,0.5)--(1,0.5)--(1,-1.5);
			\draw[red,thick,draw opacity=0.5](3.5,0.5)--(3.5,-1.5)--(10.5,-1.5)--(10.5,0.5)--(1,0.5)--(1,-1.5);
			\draw[red,draw opacity=0.5](3.5,0.5)--(3.5,-1.5)--(10.5,-1.5)--(10.5,0.5)--(1,0.5)--(1,-1.5);
			
			\node[align=center] at (1,-1.8) {$\lambda$=780nm};
			\draw[->,red,ultra thick,draw opacity=0.5](8.4,0.5)--(8.5,0.5);
			\draw[->,red,ultra thick,draw opacity=0.5](8.4,-1.5)--(8.5,-1.5);
			\draw[->,red,ultra thick,draw opacity=0.5](1,1.9)--(1,2);
			\draw[->,red,ultra thick,draw opacity=0.5](2.4,0.5)--(2.5,0.5);

			\filldraw[draw=black,fill=glass](3.2,-1.2)--++(0.6,-0.6)--++(-0.1,-0.1)--++(-0.6,0.6)--cycle;
			\node at (2.9,-1.8) {mirror};
			
			\filldraw[draw=black,fill=glass](10.3,0.9)--++(0.6,-0.6)--++(-0.1,-0.1)--++(-0.6,0.6)--cycle;
			\node at (9.6,-1.8) {mirror};
			
			\filldraw[draw=black,fill=glass](10.8,-1.2)--++(-0.6,-0.6)--++(0.1,-0.1)--++(0.6,0.6)--cycle;
			\node at (9.6,0.85) {mirror};
			
			\begin{scope}[shift={(6.5,1.5)}]			
				\filldraw[draw=black,fill=purple1](0,0)rectangle+(1,1);
				\draw[ultra thick,rounded corners=0.25cm,dashed](0,0.5)--(0.5,0.5)--(0.5,0);
			\node at (0.5,1.2) {rf switch};	
			\end{scope}
			
			\begin{scope}[shift={(5.5,3)}]			
				\filldraw[draw=black,fill=purple1](0,0)rectangle+(1,1);
				\draw[ultra thick,rounded corners=0.25cm,dashed](0,0.5)--(0.5,0.5)--(0.5,0);
			\node at (0.5,1.2) {rf switch};
			\end{scope}
			
			\filldraw[draw=black,fill=green1](0,4.5)rectangle+(2,2);
			\node[align=center] at (1,5.5) {wavemeter\\ $\delta$=150GHz};
			
			\filldraw[draw=black,fill=orange1](2.5,4.5)rectangle+(2,2);
			\node[align=center] at (3.5,5.5) {arbitrary\\ function\\ generator};
			
			\begin{scope}[shift={(9,4.5)}]
				\filldraw[draw=black,fill=pink1](0,0)rectangle+(2,2);
				\draw(0,1.33)--(0.33,1.5)--(0,1.66);
				\node[align=center] at (1,1) {pulse\\ generator};
			\end{scope}

			\begin{scope}[shift={(6.5,4.5)}]
				\filldraw[draw=black,fill=gold1](0,0)rectangle+(1.236,2);
				\node at (0.25,0.5) {E};
				\node at (0.25,1.5) {D};
				\node at (0.95,1.5) {Q};
				\node at (0.618,2.2) {D flip-flop};
			\end{scope}			
			
			\draw[ultra thick](4.5,5)--(6.5,5);

			\filldraw[draw=black,fill=theGrey](5.5,0)rectangle+(1,1);
			\node at (6,0.5) {AOM};
			\filldraw[draw=black,fill=theGrey](5.5,-2)rectangle+(1,1);
			\node at (6,-1.5) {AOM};		
				
			\filldraw[red](10.5,-0.5) circle(0.309);
			\node at (9.6,-0.5){atoms};
			
			\begin{scope}[shift={(9.675,3.6)},scale={0.5}]
				\draw(0.05,0)
				--(0.22,0)--(0.22,0.5)--(0.28,0.5)--(0.28,0)
				--(0.62,0)--(0.62,0.5)--(0.68,0.5)--(0.68,0)
				--(1.02,0)--(1.02,0.5)--(1.08,0.5)--(1.08,0)
				--(1.25,0);
				\draw(0.65,0.25) circle(0.65);
				\draw[->,rounded corners = 0.25cm](0.65,-0.4)--(0.65,-1)--(0,-1);
			\end{scope}
			
			\begin{scope}[shift={(4.8,5.35)},scale={0.5}]
				\draw(0.05,0)
				--(0.15,0)--(0.15,0.5)--(0.35,0.5)--(0.35,0)
				--(0.55,0)--(0.55,0.5)--(0.75,0.5)--(0.75,0)
				--(0.95,0)--(0.95,0.5)--(1.15,0.5)--(1.15,0)
				--(1.25,0);
				\draw(0.65,0.25) circle(0.65);
				\draw[->,rounded corners = 0.25cm](1.3,0.25)--(1.9,0.25);
				\node at (1.7,-0.25) {$\omega_p$};
			\end{scope}
			
			\begin{scope}[shift={(3.175,3.6)},scale={0.5}]
				\begin{scope}[shift={(0.05,0)}]
					\draw(0,-0.2)
					cos(0.1,0) sin(0.2,0.2)
					cos(0.3,0) sin(0.4,-0.2)
					cos(0.5,0) sin(0.6,0.2)
					cos(0.7,0) sin(0.8,-0.2)
					cos(0.9,0) sin(1,0.2)
					cos(1.1,0) sin(1.2,-0.2);
				\end{scope}
				\draw(0.65,0) circle(0.65);
				\draw[->,rounded corners = 0.25cm](0.65,-0.65)--(0.65,-1.25)--(1.5,-1.25);
		
			\end{scope}
			
			
		\end{tikzpicture}